\documentclass[11pt]{article}

\usepackage{epsfig}
\def\al{\alpha^{\prime}}
\def\a{& \hspace{-7pt}}

\def\bea{\begin{eqnarray}}
\def\eea{\end{eqnarray}}
\def\beq{\begin{equation}}
\def\eeq{\end{equation}}

\def\be{\begin{equation}}
\def\ee{\end{equation}}
\def\nn{\nonumber}
\def\a{& \hspace{-7pt}}

\def\5{\overline 5}
\newcommand{\gsim}{\mbox{\raisebox{-1.ex}{$\stackrel
     {\textstyle>}{\textstyle\sim}$}}}

\newcommand{\square}{\kern1pt\vbox{\hrule height
1.2pt\hbox{\vrule width 1.2pt\hskip 3pt
   \vbox{\vskip 6pt}\hskip 3pt\vrule width 0.6pt}\hrule
height 0.6pt}\kern1pt}
\begin{document}

\thispagestyle{empty}

\begin{center}
%\hfill ?????    \\
\hfill SISSA-3/2003/EP \\

\begin{center}

\vspace{1.7cm}

{\LARGE\bf Aspects of String-Gas Cosmology \\[2mm] at Finite Temperature}

\end{center}

\vspace{1.4cm}

{\sc B.A.  Bassett$^{a}$, M. Borunda$^{b}$, M. Serone$^{b}$
and S. Tsujikawa$^{c}$ }\\

\vspace{1.2cm}

${}^a$
{\em Institute of Cosmology and Gravitation, University of Portsmouth,} \\
{\em Mercantile House, Portsmouth PO1 2EG, United Kingdom}

\vspace{.3cm}

${}^b$
{\em ISAS-SISSA, Via Beirut 2-4, I-34013 Trieste, Italy} \\
{\em INFN, sez. di Trieste, Italy}
\vspace{.3cm}

${}^c$
{\em Research Center for the Early Universe, University of Tokyo,} \\
{\em Hongo, Bunkyo-ku, Tokyo 113-0033, Japan}

\vspace{.3cm}

\end{center}

\vspace{0.8cm}

\centerline{\bf Abstract}
\vspace{2 mm}
\begin{quote}\small
We study string-gas cosmology in dilaton gravity, inspired by the fact
that it naturally arises in a string theory context.  Our main interest is the
thermodynamical treatment of the string-gas and the resulting implications for 
the cosmology.  Within an adiabatic approximation, thermodynamical 
equilibrium and a small, toroidal universe as initial conditions, 
we numerically solve the corresponding equations of motions in 
two different regimes describing the 
string-gas thermodynamics: (i) the Hagedorn
regime, with a single scale factor, and (ii) an almost-radiation 
dominated regime, which includes the leading corrections due to the 
lightest Kaluza Klein and winding modes, with two scale factors.  
The scale factor in the Hagedorn
regime exhibits very slow time evolution with 
nearly constant energy and
negligible pressure. By contrast, in case (ii) we find interesting cosmological
solutions where the large dimensions continue to expand and the small ones
are kept undetectably small.

\end{quote}

\vfill

%%%%%%%%%%%%%%%%%%%%%%%%%%%%%%%%%%%
\section{Introduction}
%%%%%%%%%%%%%%%%%%%%%%%%%%%%%%%%%%%

String cosmology is rapidly growing in importance for at least two
reasons. First, unless there  is a fortunate conspiracy of scales and
one or more of the extra dimensions are $\sim O({\rm Tev}^{-1})$ \cite{large},
we cannot reasonably expect to detect any stringy effects in accelerators
in the foreseable future.  Hence string
cosmology may provide the only means of testing string or
M-theory concretely, through {\it e.g.} non-commutative effects \cite{non_com} 
or the AdS/CFT correspondence \cite{mald}.

On the other hand, traditional cosmology has been fettered by the chains
of the singularity theorems of General Relativity and is therefore usually
only seen as understanding the evolution of our universe over the past 15
billion years or so since the Big Bang. String theory offers the 
exciting possibility of a
resolution to the big bang singularity thereby opening up a potentially
infinite pre-history of the big-bang. Models constructed in this vein
include the pre-big-bang \cite{Veneziano:1991ek,pbb} (see
also \cite{Lidsey:1999mc,Gasperini:2002}) and the recent ekpyrotic/cyclic
models \cite{ekpyr,cyclic} which has lead to new work on string propagation
in orbifold backgrounds with curvature singularities \cite{orbsing}.

Traditional cosmology, because of the limitations of General Relativity,
suffers from another great lacking: it is unable to make any predictions
about the number of dimensions we live in or about
the spatial topology of the Universe.
String and M-theory theory, in contrast, predict that we live in 
either 10 or 11 spacetime dimensions.
Perhaps the greatest challenge for string-cosmology, after
understanding the big-bang singularity, is to explain why and how
three dimensions became observable and large while 6 or 7 are either small or
unobservable for some other reason. Perhaps the only proposal in this direction
so far is the so-called Brandenberger-Vafa (BV) scenario \cite{BV} where 
it is assumed that the universe is small
and compact and that exactly three space dimensions become large
because of the dynamics of winding modes, which play a particularly
important role.  Subsequently it has been pointed out in \cite{Veneziano:1991ek}
that the low energy effective action of a dilaton-gravity system, naturally emerging
in string theory, has a duality symmetry that is a manifestation of the
string T-duality $R\rightarrow \al/R$ symmetry, that plays a crucial role
in the analysis of \cite{BV}.  In this respect, the dilaton-gravity system
is more suitable than standard General Relativity for the BV scenario 
\cite{BV}.  Although some works have already shown that the BV scenario
can  be realized and extended in a variety of ways
\cite{TV,Tse,Miguel,CR,Sakella,Alex,Brand,Easson,EGJ,Watson,Easther} 
(see also \cite{Leblanc}), a full string-theory analysis is very 
complicated and still not completed, even in the 
simplest toroidal compactification.

In this paper, inspired by the BV scenario, we numerically solve the 
dilaton-gravity equations of motion \cite{TV} with some types 
of ``stringy matter''. Adiabatic evolution (which implies constant entropy), 
weak string coupling and
thermodynamical equilibrium are always assumed in our analysis. 
For simplicity, we analyze
Type IIA or IIB closed string theory on a $T^9$ torus, with no
branes\footnote{In \cite{Alex}
it has been shown that fundamental strings, even in the presence of D-branes,
are still the dominant degrees of freedom for the realization of the BV scenario.}.
In particular, we consider the following two regimes:
\begin{itemize}
\item (i) Hagedorn matter at high energy densities in a very small homogeneous and isotropic
universe with a common compactification radius $\sim \sqrt{\al}$, and
\item (ii) an almost-radiation dominated regime with two independent scale factors,
associated with the large and small dimensions.
\end{itemize}
In the latter case, the lightest Kaluza Klein (KK) and winding mode
contributions are also taken into account. 

It is important to stress here that in both of these regimes, our matter is manifestly
$T$-duality invariant. This symmetry is broken in our set-up only by the (arbitrary)
choice of initial values.
The main relevant questions in the two cases are respectively:
which is the evolution of the universe at early times for a free Hagedorn
string gas in thermal equilibrium?  Assuming large and small dimensions as
initial conditions, how do they evolve? In particular,
do the small dimensions remain small?

String matter in the Hagedorn phase was already briefly discussed in \cite{TV}
where it was realized that, to a first approximation in which the energy is 
constant, it leads to a very slow evolution of the universe, as in eq.~(\ref{anahagedorn}).
We complete the analysis of \cite{TV} by relaxing this approximation 
and imposing the conservation laws for KK and winding modes. We find that 
in any practical sense there is no departure from the behaviour dictated by 
eq.~(\ref{anahagedorn}), and thus no relevant dynamics emerges in this set-up.
It is important to recall how to interpret this result and its
connection to the original BV scenario.
The above result is obtained by assuming thermal equilibrium and a free, ideal, string
gas, whereas the dynamics and interactions of winding modes at very early times
are crucial in the BV scenario.
The results we get in case (i) are therefore not closely connected with the original
BV proposal \cite{BV} or, rather, they have to be taken the other way around.
Namely, unless string interactions are taken into account and/or thermal equilibrium
is relaxed, no interesting dynamics emerges.

On the other hand, the evolution of the small and large
dimensions in case (ii) is much more interesting.
First of all, we will show that when there is only pure radiation, 
the small dimensions can be stabilized and kept small 
relative to the large dimensions. Essentially, it is only required that the 
initial expansion rate of the large dimensions is bigger or of the same 
order than that of the small ones\footnote{It is actually quite hard to 
imagine initial conditions where the expansion rate of the small dimensions 
are bigger than that of the large ones.}.
This mainly comes from the fact that the pressure in the small dimensions 
vanishes in the case of pure radiation.

When matter, in the form of KK and winding modes, is included, 
the choice of initial conditions becomes more relevant.  
The crucial point is played by winding modes that are 
able to distinguish large and small dimensions, leading respectively to a 
positive/negative contribution to the pressure along the large/small 
dimensions, as is clear from eqs.~(\ref{PdM}) and (\ref{PdM2}).  This turns 
out in an expansion of the large dimensions and at the same time the 
possibility of keeping almost constant the small ones.
In fact we found that there exists a wide range of parameters for which the 
small dimensions actually remain small (see fig.~\ref{small}), while the 
large ones expand as required in the presence of radiation and string 
matter (see fig.~\ref{large}).  This is actually achieved for the natural 
initial condition  where the small dimensions are close to the 
the self-dual radius $\sqrt{\al}$, together with the condition of the expansion rate 
mentioned in the pure radiation case.

These are the most important results of the paper.  There 
exist a wide range of initial conditions for which the small dimensions are 
stabilized around the self-dual radius before entering into a purely 
radiation dominated phase, regime in which they are asymptotically stabilized 
anyway to a nearly constant value.

There are clearly several open issues that we do not consider but that deserve
further investigation. First of all,
thermal equilibrium, as well as isotropy and homogeneity
along all directions, should be relaxed and string interactions need to be
taken into account.  Further, since toroidal compactifications are not realistic,
always leading to non-chiral theories, study of more general topologies is required.
See \cite{Easson,EGJ} for recent progress in this direction.

Another very important issue unexplored in this work is the transition from
dilaton gravity to General Relativity.  A common lore is to assume that
supersymmetry breaking occurs at later times and stabilizes
all moduli, including the dilaton, which then becomes massive\footnote{An
alternative would be the models in which the dilaton stays massless but has
universal couplings.  It is even possible to reconcile this with
quintessence if the dilaton has a divergent runaway behaviour
\cite{DP1,run}.}.  It is not clear how the evolution of the large and small
dimensions are altered by this transition to an effective description in terms of 
General Relativity. 
Notice, however, that if one considers in case (ii) pure radiation with no matter,
then the late-time behaviour of the dilatonic gravity solutions lead to a standard
Friedmann-Robertson-Walker (FRW) radiation-dominated cosmology for the large dimensions,
whereas the small ones are constant. This behaviour nicely matches with the solution found
by \cite{RDSS} in General Relativity in a purely KK extra-dimensional scenario.

Finally, we assume an adiabatic approximation to simplify the 
analysis. A deeper justification of this approximation, or the dynamics which 
arises when it is dropped, is required. One can naively expect dropping this 
condition to yield a wider variety of dynamics, including more violent evolution.  

This paper is organized as follows. In section 2 some general aspects of
dilaton gravity are briefly reviewed.  In section 3 we analyse
the dynamics of the system in the extreme Hagedorn regime of high energy
densities with a single scale factor. In section 4 the dilaton-gravity equations
are solved in an almost-radiation dominated regime with large and small
scale factors, as well as in the presence of some massive stringy matter. 
 We present some
conclusions in the final section and present some useful formulae in the
appendix.  We set the string scale $\al=1$ in the following.

%%%%%%%%%%%%%%%%%%%%%%%%%%%%%%%%%%%
\section{Dilaton gravity}
%%%%%%%%%%%%%%%%%%%%%%%%%%%%%%%%%%%

We shall study the dilaton gravity equations of motion with a
massless dilaton field $\Phi$ corresponding to the low-energy effective action
of string theory in $D+1$ space-time dimensions, described by 
 \cite{Veneziano:1991ek,pbb,TV}:
\begin{eqnarray}
S=\int d^{D+1} x \sqrt{-g} \left[ e^{-2\Phi}
\left\{R+4(\nabla \Phi)^2 \right\}+{\cal L}_M \right],
\label{Baction}
\end{eqnarray}
where $g$ is the determinant of the background metric $g_{\mu \nu}$,
and ${\cal L}_M$ corresponds to the Lagrangian of some matter.
The coupling of $\Phi$ with gravity is the standard one arising in string theory.
Hereafter we shall consider the case with $D=9$.
We are interested in the case where the whole universe is small and
compact, corresponding to a flat $T^9$-torus. In this case, if one considers field configurations that are
spatially homogeneous, assuming only slow time-dependence and adopting an adiabatic
approximation it has been shown in \cite{Veneziano:1991ek,TV} that the action (\ref{Baction})
exhibits a duality symmetry, a low energy manifestation of the string $T$-duality
$R\rightarrow 1/R$ symmetry.
The ansatz for the metric and dilaton we use is
\bea
   ds^2 = -dt^2+\sum_{i=1}^9 R^2_i(t) dx_i^2\,,~~~
   R_i = e^{\lambda_i(t)}\,,~~~\Phi = \Phi(t)\,.
\label{metric}
\eea
Here the $R_i$ denotes the $i$-th scale-factor of the torus. The equations of motion 
simplify if one introduces  a shifted dilaton, $\psi$, via
\be
\psi \equiv 2 \Phi - \sum_{i=1}^9 \lambda_i\,.
\label{psidef}
\ee
Given the metric (\ref{metric}) the equations of motions of the dilaton-gravity system
are then \cite{TV}
\begin{eqnarray}
    -\sum_{i=1}^9 \dot{\lambda}_i^2 +\dot{\psi}^2 &=& e^{\psi} E,
    \label{TVb1} \\
   \ddot{\lambda}_i - \dot{\psi} \dot{\lambda}_i &=& {1\over 2} e^{\psi} P_i,
   \label{TVb2} \\
   \ddot{\psi} - \sum_{i=1}^9 \dot{\lambda}_i^2 &=& {1\over 2} e^{\psi} E,
   \label{TVb3}
\end{eqnarray}
with $E$ the total energy and $P_i$ the total pressure along the
$i$-th direction found by multiplying the total spatial volume of the space by the 
energy density and pressure appearing in ${\cal L}_M$ of (\ref{Baction}). 
Here a dot denotes derivative with respect to cosmic
time, $t$. These equations are manifestly invariant under the duality symmetry
\cite{Veneziano:1991ek,TV}
\be \lambda_i \rightarrow - \lambda_i, \ \ \ \Phi \rightarrow
\Phi -\sum_i \lambda_i,
\label{dualsym}
\ee
under which $\psi$ defined in eq.~(\ref{psidef}) is left invariant.  It is
typically assumed that the scale factors $R_i$ are the same in all
directions, {\it i.e.} $R_i=R$.  In contrast, we also consider in section 4 a scenario
where the background is homogeneous and isotropic in $d$-spatial large
dimensions and $(9-d)$-spatial small dimensions.  We denote the large and
small dimensions with their corresponding scale factors, as
\bea
R = e^{\mu} \,,~~~r = e^\nu\,.
\eea
In this case eqs.~(\ref{TVb1})-(\ref{TVb3}) take the form
\begin{eqnarray}
   -d \dot{\mu}^2 - (9-d) \dot{\nu}^2 + \dot{\psi}^2 &=& e^{\psi} E,
   \label{b1}  \\
   \ddot{\mu} - \dot{\psi} \dot{\mu} &=& {1\over 2} e^{\psi} P_d,
   \label{b2} \\
   \ddot{\nu} - \dot{\psi} \dot{\nu} &=& {1\over 2} e^{\psi} P_{9-d},
   \label{b3} \\
   \ddot{\psi} - d \dot{\mu}^2 - (9-d) \dot{\nu}^2  &=& {1\over 2} e^{\psi} E,
   \label{b4}
\end{eqnarray}
where
\bea P_d &=&
-\frac{\partial F}{\partial \mu_i},~~~~\ \ \forall i=1,\ldots, d \,, \\
P_{9-d} & = & -\frac{\partial F}{\partial \nu_i},~~~~\ \ \forall i=d+1,\ldots, 9 \,,
\eea
in terms of the free energy $F$.

%%%%%%%%%%%%%%%
\section{Hagedorn regime}
%%%%%%%%%%%%%%%

According to the original BV proposal \cite{BV}, the very early universe was compact
over all nine spatial dimensions with radii
$r\sim 1$ in string units. In this section we study the dilaton-gravity equations
(\ref{TVb1})-({\ref{TVb3}), with $E$ and $P$ the total energy and pressure of a
free string gas in thermal equilibrium in such a compact, small universe. For simplicity,
we consider the string gas associated with type IIA/IIB string theory
compactified on a square $T^9$-torus, simple product of nine circles, with
radii all equal to a common value $r=e^{\lambda}$.  Although the strict
thermodynamical limit $V\rightarrow \infty$ cannot be taken for this
system, thermodynamics is still trustable as long as the system contains
many degrees of freedom.  In our case, this implies having an energy
density $\rho\gg 1$.  Our first step is then to derive the equation of
state of the string matter in this regime, or equivalently the energy and
pressure entering in eqs.~(\ref{TVb1})-(\ref{TVb3}).

\subsection{Microcanonical ensemble}

String thermodynamics in the high density phase has been a subject of intense
study in the past (for a review see {\it e.g.}
\cite{Alvarez-Gaume:1992re}).  The most important point is the emergence of
a critical temperature, the Hagedorn temperature $T_H$ \cite{Hage}, where
the partition function of a free string gas diverges.  It was soon 
realized that in this regime the usual thermodynamical equivalence between
the canonical and microcanonical ensembles can break down and the latter,
more fundamental ensemble, must hence be used.
Furthermore, due to the
presence of winding modes, the thermodynamical properties of a string
system at finite volume differs substantially from that of an infinitely
extended system \cite{BV,DJT}.  Following \cite{DJT}, it is useful to
derive the energy density of states $\Omega(E)$ by analyzing the
singularities of the one-loop string partition function in the complex
$\beta$-plane, where $\beta=1/T$ is the inverse temperature.  This analysis
has already been carried out in \cite{DJT} but for completeness we report some
useful details in the Appendix.

For a totally compact space at high energy, the leading singularity (a simple pole)
of the partition function at $\beta=\beta_H$
is not sufficient to establish the thermodynamical properties of the system \cite{BV,DJT}.
The first next-to-leading singularities
are poles of order 18 with a  dependence on the compactification radius.
In this case a useful parametrization of the one-loop partition function $Z$ is given in
eq.~(\ref{ZHage}), from which one computes the density of states $\Omega(E)$ by means of 
eq.(\ref{omega}) and thus the associated entropy $S=\log \Omega(E)$ 
\footnote{We have numerically checked that the term
$\Lambda(\beta, R)$ in eq.~(\ref{ZHage}) is negligible
and thus is not reported in the following.}.
It reads
\be
S(E,r) \simeq \beta_H E+
\log\Bigg\{ 1- \frac{1}{\Gamma(18)\eta_{KW}^{18}}\bigg[(\eta_K E)^{17} \eta_W^{18} e^{-\eta_K E}
+ (\eta_W E)^{17} \eta_K^{18} e^{-\eta_W E} \bigg]\Bigg\}\,,
\label{Omega0d0}
\ee
where $\beta_H=2\sqrt{2}\pi $, and
\bea
\label{etas}
\eta_K \a = \a \sqrt{2} \pi  \bigg[2 - \sqrt{4-\frac{2}{r^2}}\bigg]\,,\nn \\
\eta_W \a = \a \sqrt{2}\pi  \bigg[2 - \sqrt{4-2r^2}\bigg]\,, \\
\eta_{KW} \a = \a \sqrt{2}\pi  \Bigg\{ \bigg[2 - \sqrt{4-\frac{2}{r^2}}\bigg]
- \bigg[2 - \sqrt{4-2r^2}\bigg] \Bigg\}\,. \nn
\eea
The energy as a function of $r$ is given directly by (\ref{Omega0d0}),
since $S=$ constant, by the assumption of adiabatic evolution.  On the
other hand, the temperature and pressure, defined as
\be
\frac
1T = \frac{\partial S}{\partial E} \,, \ \ \ \ \ \ \
P = \frac T9\frac{\partial S}{\partial (\log r)}\,,
\ee
yield
\bea
\label{tem}
\frac{1}{T} &=& \frac{1}{T_H}-\frac{1}{xE\, \Gamma(18)}
\Biggl[ \biggl(\frac{z}{w}\biggr)^{18} y^{17}(17-y)e^{-y}+
 \biggl(\frac{y}{w}\biggr)^{18} z^{17}(17-z)e^{-z} \Biggr]\,,
 \\
P&=& \frac{T}{9} \frac{\dot{x}}{x} \frac{1}{\dot{\lambda}}\,,
\label{prehage}
\eea
where
\bea
\label{xhage}
x \equiv 1- \frac{1}{\Gamma(18)} \left( \frac{yz}{w}\right)^{18}
\left( \frac{e^{-y}}{y}+\frac{e^{-z}}{z} \right)\,,
\eea
and $y \equiv \eta_K E$, $z \equiv \eta_W E$,
$w \equiv \eta_{KW}E$.
As usual in the microcanonical ensemble, the temperature is a derived quantity
(from $S$ and $E$) and its explicit form is needed
only to compute the pressure $P$.
When the radius $r$ is close to unity, $T$ and $P$ are
approximately given by
\bea
\label{appT}
\frac{1}{T}\a\sim\a   \frac{1}{T_H} + C_1 E^{17}
e^{-\tilde{\eta} E}\,, \\
P \a\sim\a C_2 E^{17} e^{-\tilde{\eta} E}\,,
\label{PT}
\eea
where $\tilde{\eta} \simeq \eta_K\simeq\eta_W$ for $r\sim 1$, and $C_1$ and
$C_2$ are certain polynomial functions of $\tilde{\eta}$ and $\eta_{KW}$.

We numerically solved the dilaton-gravity equations
(\ref{TVb1})-(\ref{TVb3}) using a standard Runge-Kutta
routine. We adopted initial conditions around $E_0\sim 1000$, $r_0 \sim
1$, which comes from the requirement of T-duality.
The shifted dilaton is chosen to satisfy the condition
$e^{\Phi} \ll 1$ to ensure that the string coupling
constant is initially small and hence that perturbation theory and the ideal gas 
approximation are trustable.  The initial
condition for $\dot\lambda$ is somewhat arbitrary and we have carried out
simulations for a wide variety of different initial values of $\dot\lambda$.  Notice that
$\dot\psi_0^2$ is fixed by the constraint equation (\ref{TVb1}) and that
the negative solution is taken to remain in the perturbative regime
of small string coupling constant.

For initial conditions $r_0\in [0.8,1.2]$ and $E_0\sim 1000$,
the temperature is very close to the Hagedorn temperature with a nearly constant
value.  This is clear from eq.~(\ref{appT}), since the last term in (\ref{appT})
is vanishingly small relative to the first term, due to the exponential
suppression given by $e^{-\tilde{\eta} E}$.  Similarly we have
$P \simeq 0$ for the above initial conditions from eq.~(\ref{PT}).
Therefore the system is effectively described by a pressureless dust
as shown in fig.~\ref{hagedorn}.
In this case one has $\dot{\lambda} \simeq A
e^{\psi}$ from eq.~(\ref{TVb2}), with $A$ an integration constant.
Subtracting eq.~(\ref{TVb3}) from eq.~(\ref{TVb1}), we find a simple
relation, $\ddot{(e^{-\psi})}=E/2$.  Taking note that $E$ is nearly
constant ($E \simeq E_0$), the analytic solutions of
eq.~(\ref{TVb1})-(\ref{TVb3}) in the Hagedorn regime may be written as
\begin{eqnarray}
   e^{-\psi} &\simeq&  \frac{E_0}{4}t^2+Bt+\frac{B^2-dA^2}{E_0}\,, \\
  \lambda &\simeq& \lambda_0+\frac{1}{\sqrt{d}}
  {\rm log} \left| \frac{(E_0 t+2B-2\sqrt{d}A)(B+\sqrt{d}A)} {(E_0
  t+2B+2\sqrt{d}A)(B-\sqrt{d}A)} \right|\,.
   \label{anahagedorn}
\end{eqnarray}
$A$ and $B$ are integration constants depending on the initial values for $\dot\lambda$,
$\psi$ and $\dot \psi$. In particular
\be
A = \dot\lambda_0 e^{-\psi_0}, \ \ \ \  B=-\dot\psi_0 e^{-\psi_0}\,,
\ee
and $d$ is the number of dimensions (we are now considering the case
with $d=D=9$). Notice that due to eq.~(\ref{TVb1}), $\dot\psi_0$, and thus $B$,
can not be taken to be vanishing.

In fig.~\ref{hagedorn} we plot the evolution of $r$ that
corresponds to the analytic solution (\ref{anahagedorn}), together with the
full numerical results.  They show very good agreement, which implies that
the Hagedorn regime is well described by a state with a constant energy
and negligible pressure.  This actually ensures the validity of the
analytic estimation in ref.~\cite{TV} discussed briefly in its Appendix.

As long as $\dot\lambda_0$ is positive (negative), the radius grows
(decreases) towards the asymptotic value
\begin{eqnarray}
 r_{\infty}=e^{\lambda_0}
 \left| \frac{B+\sqrt{d}A}{B-\sqrt{d}A} \right|^{1/\sqrt{d}} \,,
\end{eqnarray}
with $\dot{r}$ getting smaller with time (see fig.~\ref{hagedorn}).
We have checked this for values of $r_0$ very
close to $1$, up to $r_0=(1+1\times 10^{-15})$, and found no substantial
changes in behaviour.  If one chooses exactly $r_0=1$,
eq.~(\ref{Omega0d0}) should be replaced by another similar relation, since
now the two poles of order 18 in the $\beta$-plane approach each other to a
single pole of order 36.  The above results apply also in this case: the
pressure is almost zero and the evolution of the system is very slow in
time.

For initial values of $E$ in the range $E_0 \in [500,5000]$,
the dynamics of the system is practically the same as explained above.
For initial energies $E_0~\gsim~1000$ the scale factor is essentially constant in
time.  On the other hand the system is typically unstable for $E_0 \ll 1000$ and
not thermodynamically meaningful for such low values of $E_0$.

%%%
\begin{figure}
\epsfxsize = 4.5in
\epsffile{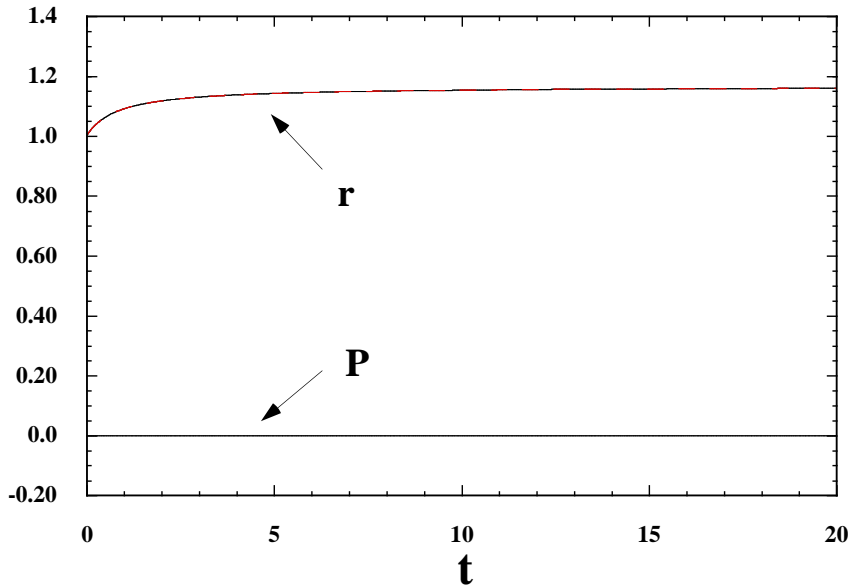}
\caption{The evolution of $r$ and $P$ for the initial conditions
$\lambda_0=10^{-3}$, $\dot{\lambda}_0=0.2$, $\psi_0=-5$ and
$E_0=10^3$ in the Hagedorn regime.
$\dot{\psi}_0$ is determined by the constraint
equation (\ref{TVb1}).  We plot the evolution of $r$ both using the 
analytic approximation (\ref{anahagedorn}) and the full
numerical result, which show very good agreement each other.}
\label{hagedorn}
\end{figure}
%%%

%
\subsection{Conservation laws}

We can also address the question of what happens when conservation laws
are taken into account.  In the case of the toroidal compactification we
consider, the conserved quantities are taken to be the total winding number
$N_i$ and the KK momenta $M_i$ in each compact dimension ($i=1,\dots , 9$).
This is performed by introducing a chemical potential for each conserved
charge \cite{DJT2}.

A crucial difference with respect
to the previous case is that the leading singularity of the partition
function, namely the Hagedorn temperature, is enough to study
the thermal behaviour of the system. 
In particular, the leading singularity now   
depends on the compactification radius induced by the conservation laws.
The entropy of such an ideal gas can be easily computed, yielding
\be
S(E,r) \simeq
\beta_H E\, -\, \frac{\pi}{4\sqrt 2\,E} \sum_{i=1}^9\left
(\frac{M_i^2}{r^2}+N_i^2 r^2 \right ) -9\,\ln E\,.
\label{conhag}
\ee
We see that in eq.~(\ref{conhag}) there are two suppression
terms with respect to the leading term $\beta_H E$ in
eq.~(\ref{Omega0d0}). This is expected, because eq.~(\ref{Omega0d0}) counts states
with all values of charges whereas (\ref{conhag}) counts a smaller set of
states in which the value of the charge is fixed.  Moreover, the number of states
decreases as $M_i$ or $N_i$ are increased, since less energy is
available for the oscillators.  The temperature and pressure obtained
from the entropy (\ref{conhag}) are:
\bea
\label{sum2}
\frac 1 T &=& \frac{1}{T_H} -\frac 9 E +\frac{1}{2\sqrt 2
\pi}\frac 1 {E^2} \sum_{i=1}^9\left( \frac{M_i^2}{r^2}+ N_i^2 r^2\right)\,,
\\
P &=& \frac T 9 \frac{\pi}{2\sqrt2} \frac 1 E\sum_{i=1}^9
\left(\frac{M_i^2}{r^2}-N_i^2r^2 \right)\,.
\label{sum3}
\eea
The pressure vanishes if one imposes  vanishing winding and KK
charge, $M_i=N_i=0$.
The energy $E$ evolves very slowly for $E_0$ of order 1000 
in which case the radius $r$ asymptotically approaches a constant value after
some growth from $r_0 \simeq 1$, thereby showing similar behaviour to
fig.~\ref{hagedorn}.  When $M_i,N_i\neq 0$, due to presence of the $E$
factor in the denominator of eq.~(\ref{sum3}), the system evolves similarly
to the case $M_i=N_i=0$, as long as the summation terms in
eqs.~(\ref{TVb2}) and (\ref{TVb3}) are unimportant.  The dynamics changes
if $M_i$ and $N_i$ are of order $10^5$, since the scale factor can have
respectively a significant expansion or decreasing rate.
As expected, winding modes prevent expansion, whereas KK modes, as standard
matter, favorite it.

As a last remark, notice that the string gas defined by eqs.~(\ref{sum2})
and (\ref{sum3}) has a negative specific heat.
Systems with negative specific heat are unstable
in non-compact spaces, but actually can be in thermal equilibrium in a finite
space. Along the lines of \cite{hawking,AEK}, we have evaluated the critical
volume $V_c$ under which the system is in equilibrium and found that this is
actually the case for $V_c\gg 1$, implying that the system can be actually in
equilibrium with radiation.  It should also be emphasized that this system
is trustable as long as string corrections are completely neglected.  When
string interactions are included, the system is most likely to undergo a
phase transition \cite{AW} whose details are so far unknown.

%%%%%%%%%%%%%%%%%%%%%%%%%%%%%%%%%%%
\section{Almost-radiation regime}
%%%%%%%%%%%%%%%%%%%%%%%%%%%%%%%%%%%

Assuming that some dimensions ($d$) start to expand
while the remaining $(9-d)$ dimensions remain small by means of some
mechanism, the system will eventually reach a temperature below
the Hagedorn regime where the dynamics is mostly governed by massless
states, {\it i.e.} radiation\footnote{Notice that the system is already in an almost
radiation regime for $\beta~\gsim~11$ and it is essentially governed by 
 pure radiation only for $\beta~\gsim~14$, in string units.}.  In this case, it is 
important to see the conditions under which the large dimensions continue 
to expand.  At the same time, it is important to study whether the small 
dimensions remain small or if they also enter an expanding phase.  
Different from the Hagedorn regime discussed in the section 3, the 
microcanonical ensemble agrees with the canonical one for the range of 
temperatures and energy densities involved now.  In the following we shall 
use the latter ensemble, which is more convenient for practical purposes.

Assuming again thermodynamical equilibrium and adiabatic evolution,
we shall solve the dilaton-gravity equations (\ref{b1})-(\ref{b4}),
with $E$ and $P$ obtained from the free energy of a string gas at
temperatures below $T_H$.
The entropy $S$ is conserved with time
under the assumption of adiabatic evolution:
\beq
\frac{d}{dt} S = \frac{d}{dt} \left( \beta^2 \frac{\partial F}{\partial
\beta}\right)  = 0\,.
\label{adiab}
\eeq
Eq.~(\ref{adiab}) is solved by letting $\beta$ and the scale factors
$\lambda_i = \log \, R_i$,
be slowly varying functions of time [$\beta \rightarrow \beta(t)$,
$\lambda_i\rightarrow\lambda_i(t)$].
In this way one can derive a differential equation whose solution gives
$\beta = \beta(\lambda_i)$ with $S = $constant.  
We denote the radii of the large $d$
dimensions, taken all equal, by $R=e^{\mu}$, whereas the radii
of the $(9-d)$ small dimensions, again all equal, by
$r=e^{\nu}$.

\subsection{Pure radiation}\label{massless}

As a first step, let us consider the case of pure 
radiation (see the Appendix for the free energy in the context of the
canonical ensemble).  The energy and pressure are easily evaluated from
eq.~(\ref{freed0}) :
\be
E_{rad}^{(d)} =F_{rad}^{(d)}+\beta \frac{\partial F_{rad}^{(d)}}
{\partial \beta} = \frac{d R^d}{2\pi}D(0)^2 \Gamma \left(\frac{d+1}{2}\right)
(4\pi)^{\frac{d+1}2} \zeta(d+1) (1-2^{-(d+1)}) \beta^{-d-1}\,,
\label{E0rad}
\ee
whereas the pressure $P_{rad}^{(d)}$ for the $d$ spatial dimensions is given by
\be
P_{rad}^{(d)}= -\frac1d\frac{\partial
F_{rad}^{(d)}}{\partial ({\rm ln}\, R)} = - F_{rad}^{(d)}
=\frac{E_{rad}^{(d)}}{d} \,,
\label{rela}
\ee
which corresponds to the equation of state for radiation in $d$ spatial
dimensions.
Eq.~(\ref{E0rad}) is nothing but the $d$-dimensional generalization of the
Stefan-Boltzmann law in presence of $D(0)^2/2$ bosonic and fermionic
degrees of freedom.  Since $F_{rad}^{(d)}$ does not depend on $r$, the
pressure along the small dimensions vanish:
\be
P_{rad}^{(9-d)} =
-\frac{1}{9-d}\frac{\partial F_{rad}^{(d)}}{\partial ({\rm ln}\, r)} = 0\,.
\ee
{}From the adiabatic equation (\ref{adiab}), we easily get the
following relation
\beq
\beta=\beta_0 \frac{R}{R_0}\,,
\label{betaana}
\eeq
relating the temperature and scale factor in a radiation-dominated universe,
with $\beta_0$ and $R_0$ being initial conditions satisfying
$\beta(R_0)=\beta_0$.  In this case the dilaton-gravity equations
(\ref{b1})-(\ref{b4}) read
\begin{eqnarray}
   \ddot{\psi} &=& \frac12 d\dot{\mu}^2+\frac12 (9-d)\dot{\nu}^2+
                             \frac12 \dot{\psi}^2\,, \label{MLb1} \\
                             \ddot{\mu} &=& \dot{\psi} \dot{\mu}+
                             \frac12 e^{\psi} P_{rad}^{(d)}\,,
                             \label{MLb2} \\ \ddot{\nu}&=& \dot{\psi}
                             \dot{\nu} \,, \label{MLb3}
\end{eqnarray}
together with the constraint equation
\begin{eqnarray}
   \dot{\psi}^2=e^{\psi}E_{rad}^{(d)}+d\dot{\mu}^2+ (9-d)\dot{\nu}^2\,.
   \label{MLb4}
\end{eqnarray}
Eq.~(\ref{MLb3}) is integrated to give
\begin{eqnarray}
   \dot{\nu}=\dot{\nu}_0 e^{\psi-\psi_0} \,,
   \label{dotnu}
\end{eqnarray}
where $\dot{\nu}_0$ are $\psi_0$ are the initial values of
$\dot{\nu}$ are $\psi$. We see from
eq.~(\ref{dotnu}) that when $\dot{\nu}_0$ is positive (negative) the
expansion rate for the small dimensions is always positive (negative).  In
order to avoid unbounded growth of the dilaton towards the strongly coupled
regime ($e^{\Phi}~\gsim~1$), it is natural to consider the case with
negative $\dot{\psi}$.  In this case the absolute value of $\dot{\nu}$
decreases with time.  

In the absence of the pressure $P_{rad}^{(d)}$ in eq.~(\ref{MLb2}),
the evolution of the large dimensions is similar to that of the small ones.
In the case $\dot{\psi}_0<0$ and $\dot{\mu}_0>0$, we have
$\ddot{\mu}<0$ for $P_{rad}^{(d)}=0$ from eq.~(\ref{MLb2}).
This corresponds to the
universe with expanding large dimensions with a decreasing Hubble rate.
Numerically we found that the evolution of the system in this case is
trivial, namely the large dimensions soon approach a nearly constant value
with very small $\dot{\mu}$.

When the pressure $P_{rad}^{(d)}$ is taken into account, this works as a
positive source term in eq.~(\ref{MLb2}).  Therefore it is possible to
make the r.h.s. of eq.~(\ref{MLb2}) positive even when $\dot{\psi}<0$
and $\dot{\mu}>0$.  We have made numerical simulations with initial
conditions $R_0 \gg r_0 \sim 1$, $\dot{\psi}_0<0$, and several different
values of $\dot{\mu}_0$ and $\dot{\nu}_0$. As long as
$\dot{\mu}_0$ is positive, the large dimensions expand in the
presence of the pressure due to radiation.  The contribution of the
pressure term in eq.~(\ref{MLb2}) inhibits the rapid decrease of
$\dot{\mu}$, thereby leading to different evolution of $R$ compared to
the case with $P_{rad}^{(d)}=0$.  The expansion rate $\dot{\nu}$ for the
small dimensions is exponentially suppressed with the decrease of $\psi$
[see eq.~(\ref{dotnu})].  Therefore unless the initial value of
$|\dot{\nu}|$ is much larger than unity, the radius $r$ can stay small
around $r \sim 1$.

%%%
\begin{figure}
\epsfxsize = 4.5in \epsffile{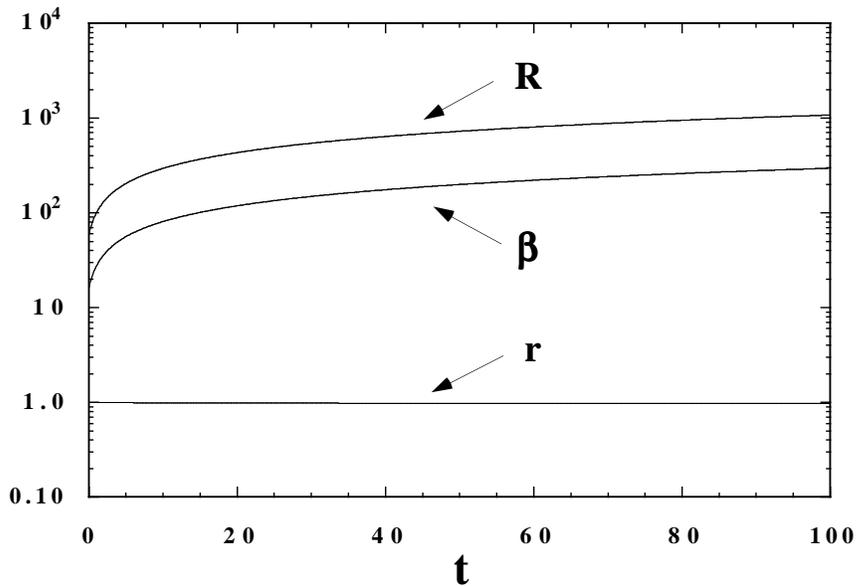}
\caption{The evolution of $R$, $r$ and $\beta$ for the pure radiation case
with $d=3$.  We choose the initial conditions $\dot{\mu}_0=1.0$, $\mu_0=4.0$,
$\dot{\nu}_0=-0.01$, $\nu_0=0.0$, $\psi_0=-16$ and $\beta_0=15$.  }
\label{mlessevo}
\end{figure}
%%%

%%%
\begin{figure}
\epsfxsize = 4.5in \epsffile{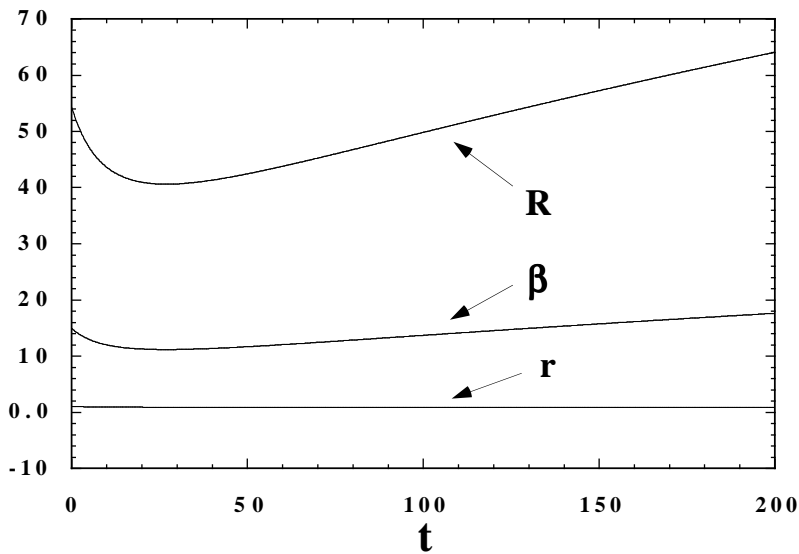} \caption{The evolution of $R$,
$r$ and $\beta$ for the pure radiation case with $d=3$.
We choose the initial conditions $\dot{\mu}_0=-0.04$, $\mu_0=4.0$,
$\dot{\nu}_0=-0.01$, $\nu_0=0.0$, $\psi_0=-16$ and $\beta_0=15$. }
\label{bounce}
\end{figure}
%%%

We have numerically succeeded to obtain ideal solutions with
growing $R$ and small, roughly constant, $r$ satisfying $r \ll R$.  
One typical evolution is 
plotted in fig.~\ref{mlessevo}.  These solutions can be achieved by 
choosing initial values with $\dot{\mu}_0~\gsim~|\dot{\nu}_0|$, 
$R_0 \gg r_0 \sim 1$ and $\dot{\psi}_0<0$.  
When $\dot{\mu}_0 \ll |\dot{\nu}_0|$ holds initially, 
it is difficult to keep the small dimensions
small relative to the large ones.
If $\dot{\mu}_0<0$, we have $\ddot{\mu}_0>0$ from 
eq.~(\ref{MLb2}).  This leads to the growth of the expansion rate 
$\dot{\mu}$.  Since $\dot{\mu}$ continues to be negative by the time it 
crosses zero, the large dimensions contract during this stage.  After 
$\dot{\mu}$ changes sign, $R$ begins to grow.  This implies that bouncing 
solutions may be obtained if $\dot{\mu}_0<0$.  We have numerically found 
that this is actually the case, see fig.~\ref{bounce}.  During the 
contracting phase, the temperature increases according to 
eq.~(\ref{betaana}).  The temperature is maximum at the bounce where $R$ is 
minimum.  In the context of Pre-Big-Bang \cite{pbb} or Ekpyrotic 
Cosmologies \cite{ekpyr}, nonsingular bouncing solutions are difficult to 
construct unless loop or derivative corrections are added to the 
tree-level action \cite{Gasperini:1996fu,Tsujikawa:2002qc}.  It is quite 
interesting to be able to obtain  bouncing solutions only by including 
radiation in dilaton-gravity equations.

It is worth investigating the asymptotic behavior of the dynamical system
of eqs.~(\ref{MLb1})-(\ref{MLb3}), along the lines of \cite{TV}.
Introducing new parameters, $\dot{\mu}
\equiv \xi$, $\dot{\nu} \equiv \eta$, $\dot{\psi} \equiv f$, and using the fact
that the pressure in the large dimensions is connected to the energy
$E_{rad}^{(d)}$ through eq.~(\ref{rela}), eqs.~(\ref{MLb1})-(\ref{MLb4})
become:
\begin{eqnarray}
  \label{reduce3}
   \dot{f} &=& \frac{d}{2} \left[ \xi^2+\left(
  \frac{9}{d}-1\right)\eta^2\right] +\frac12 f^2\,, \\
  \label{reduce}
  \dot{\xi} &=& -\frac12 \left[ \xi^2+\left(
  \frac{9}{d}-1\right)\eta^2\right] +f\xi+\frac{f^2}{2d}\,, \\
  \label{reduce2}
  \dot{\eta} &=& f\eta\,.
\end{eqnarray}
Since we are considering the case with decreasing $\dot{\psi}$,
$\eta$ asymptotically approaches zero from 
eq.~(\ref{dotnu}), {\it i.e.} $\eta=0$ is an attractive solution.  In this case, the analysis 
is closely related with the one  outlined in the Appendix of 
\cite{TV}.  In particular one finds that the line described by 
\begin{eqnarray}
  f/\xi=-d\,,~~~\eta=0\,,
  \label{att}
\end{eqnarray}
is an attractor. {}From eq.~(\ref{psidef}) the time-derivative of the
dilaton is given as $2\dot{\Phi}=f+d\xi+(9-d)\eta$.  Therefore we have
$\dot{\Phi}=0$ for the attractor (\ref{att}), again in complete analogy
to the case of the single 
scale factor \cite{TV}.  Substituting eq.~(\ref{att}) for
eq.~(\ref{reduce}) and integrating this relation, one finds
\begin{eqnarray}
\xi \propto \frac{2}{(d+1)t}\,,~~~R \propto t^{\frac{2}{d+1}}\,.
\label{radan}
\end{eqnarray}
This indicates that the late time evolution for the large dimensions can be
described by that of the standard radiation dominant phase in
FRW cosmology for $d=3$, even in the presence of the small dimensions.  The key point is
that the pressure in the small dimensions vanishes for the massless case,
thereby leading to $\eta=0$ as an attractor.
Notice that a cosmological solution of this kind has been also obtained in standard
General Relativity and in a purely Kaluza-Klein extra-dimensional scenario 
by \cite{RDSS}.

We also analyzed the evolution of the system by varying the value of
$d$, and found that the situation is not basically changed compared to
the $d=3$ case discussed above.  
As long as the initial 
conditions satisfy $\dot{\mu}_0 ~\gsim~|\dot{\nu}_0|$, 
$R_0 \gg r_0 \sim 1$ and $\dot{\psi}_0<0$,  the large 
dimensions continue to expand due to the presence of radiation 
while the small dimensions are kept to be small ($r \ll R$).
In this case the large dimensions asymptotically approach
the radiation-dominant FRW solution (\ref{radan}).
When the initial value of $\dot{\mu}$ is negative, we found that it is 
possible to have a bouncing cosmological solution that approaches the 
expanding FRW universe given by (\ref{radan}).

\subsection{Inclusion of matter} \label{massive}

Although massless states, pure radiation, dominate the thermodynamical
ensemble in this phase, this dominant contribution has a trivial
dependence on the small dimensions $r$. In particular, $F_{rad}^{(d)}$
does not depend on $r$ and the pressure along the small dimensions
trivially vanishes. It is then important to see if and how matter terms can
alter this behaviour. For this purpose, we study the leading terms that have an
explicit dependence on $r$ in the infinite sums appearing in eq.~(\ref{Fd}).
We have numerically estimated that it is enough to consider the first
KK and winding modes along a small direction, {\it i.e.} the terms
with $\{N=\bar N=0, m_i = (1,0,\ldots ,0), n_i=0\}$ (as well as $m_i$ and
$n_i$ exchanged) in eq.~(\ref{Fd}), plus
the remaining $8-d$ inequivalent permutations.  The energy $E^{(d)}_{mat}$ and 
pressures $P_{mat}^{(d)}$
and $P_{mat}^{(9-d)}$ along the large and small dimensions associated with
these states are simply evaluated starting from the general expression
(\ref{Fd}).  The equation of state for these leading order terms are:
\bea
\label{E_M2}
E_{mat}^{(d)} = \a\a -V_d \, C(\beta)^{(d)}
\Bigg\{ \bigg[ \frac{1}{r^{(d+1)/2}} \frac{1-d}{2}
K_{(d+1)/2}\bigg(\frac\beta r\bigg) + \frac{1}{r^{(d+1)/2}}
\frac{\beta}{r} K^{\prime}_{(d+1)/2} \bigg(\frac\beta r\bigg) \bigg] +
\nn \\
\a\a \hspace{2.1cm} \bigg[ r^{(d+1)/2} \frac{1-d}{2}
K_{(d+1)/2}\left(\beta r\right) + r^{(d+1)/2} {\beta}{r} K^{\prime}_{(d+1)/2}
\left(\beta r\right) \bigg]\Bigg\} \,, \\
\label{PdM}
P_{mat}^{(d)} = \a\a V_d \, C(\beta)^{(d)} \biggl[ \frac{1}{r^{(d+1)/2}}
K_{(d+1)/2}\bigg(\frac\beta r\bigg)+r^{(d+1)/2} K_{(d+1)/2} (\beta r)
\biggr]\,, \\
\label{PdM2}
P^{(9-d)}_{mat} = \a\a V_d \, C(\beta)^{(d)}
\beta \Biggl[\frac{1}{r^{(d+3)/2}} K_{(d-1)/2}\bigg(\frac\beta r\bigg)
-r^{(d+3)/2} K_{(d-1)/2} (\beta r) \Biggr]\,,
\eea
where $K_n$ are modified Bessel functions, the prime denotes derivative 
with respect to $r$, and  
\be C(\beta)^{(d)} 
=\left(\frac{2\pi}{\beta}\right)^{(d+1)/2} \frac{(18-2d)}{\pi} D(0)^2\,.
\ee 
Here $D(0)^2=256$ is a string degeneracy factor (see the Appendix).  Note 
that we only consider the $p=1$ term in eq.~(\ref{Fd}), implying the 
approximation of the bosonic/fermionic statistics with the 
Maxwell-Boltzmann distribution.  The pressure $P_{mat}^{(d)}$ along the 
large dimensions is always positive, which aids expansion 
of the universe in addition to the pressure $P_{rad}^{(d)}$ from the massless 
states.  The first and second terms in square brackets in eqs.~(\ref{PdM}) 
and (\ref{PdM2}) come from the KK and winding mode, respectively.  The 
above equations (\ref{E_M2})-(\ref{PdM2}) are all manifestly invariant 
under the duality symmetry (\ref{dualsym}) acting on the small dimensions, 
$r\rightarrow 1/r$.  Notice that the winding modes give rise to a standard 
positive pressure along the large dimensions (second term in (\ref{PdM})) 
but negative along the small ones (second term in (\ref{PdM2})).

We numerically solved the dilaton-gravity equations (\ref{b1})-(\ref{b4}),
with $E=E_{rad}^{(d)}+E_{mat}^{(d)}$,
$P^{(d)}=P_{rad}^{(d)}+P_{mat}^{(d)}$ and $P^{(9-d)}=P_{rad}^{(9-d)}
+P_{mat}^{(9-d)}$,
by carefully taking into account the adiabaticity condition (\ref{adiab}).
The pressure $P^{(9-d)}_{mat}$ for the small
dimensions vanishes at the self-dual critical radius $r=1$.
Therefore it is expected that the effect of the massive states
for the small dimensions is weak around $r \sim 1$.
In fact we have numerically found that this is the case.
As seen from the case (b) in fig.~\ref{small}, the evolution of the small
dimensions is hardly altered by including the massive mode for the initial
value of $r$ very close to unity.
{}From eq.~(\ref{PdM2}) one notes that $P_{mat}^{(9-d)}<0$ for $0<r<1$
and $P_{mat}^{(9-d)}>0$ for $r>1$ (the asymptotic values are  $P_{mat}^{(9-d)}
\to 0$ for $r \to 0$ and $r \to \infty$).  
This indicates that the pressure
of the massive state makes the small dimensions contract for $0<r<1$ while its
effect tends to expand the small dimensions for $r>1$.
%Unlike the case with pure radiation,
%we have an extra source term in the r.h.s. of eq.~(\ref{reduce2}), 
%in which case $\dot{\nu}=0$ is not necessarily an attractor.

The effect of the massive states emerges by choosing the initial values of
$r_0$ that are slightly smaller or larger than unity.  For $0<r_0<1$ with
$\dot{\nu}_0>0$, the small dimensions can be larger than $r=1$ for large
initial values of $\dot{\nu}$.
In this case the small dimensions continue to grow after they cross $r=1$.
When $\dot{\nu}_0$ is not large ($\dot{\nu}_0 \ll 1$), the massive effect
can lead to the contraction of the small dimensions due to the
negative pressure for $r<1$.
As found from the case (c) in fig.~\ref{small}, the small dimensions always
increase in the massless case, whereas the small dimensions begin to contract
if the massive effect is included.  Therefore we can keep the $(9-d)$
dimensions small ($0<r<1$) for these initial conditions.

%%%
\begin{figure}
\epsfxsize = 4.5in \epsffile{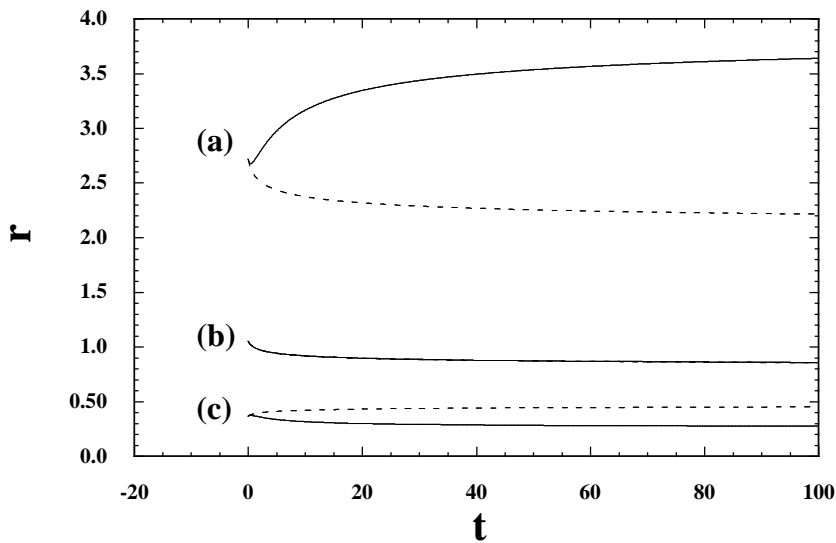}
\caption{The evolution of the small dimensions for $d=4$
when the massive states are taken into account (solid curve).
The dotted curves correspond to the case where the massive
states are neglected (only massless states).
We choose the initial conditions $\dot{\mu}_0=1.0$, $\mu_0=4.0$,
$\psi_0=-16$, $\beta_0=12$ with (a) $\dot{\nu}_0=-0.1$, $\nu_0=1.0$,
(b) $\dot{\nu}_0=-0.1$, $\nu_0=0.05$,
(c) $\dot{\nu}_0=0.1$, $\nu_0=-1.0$.}
\label{small}
\end{figure}
%%%

%%%
\begin{figure}
\epsfxsize = 4.5in
\epsffile{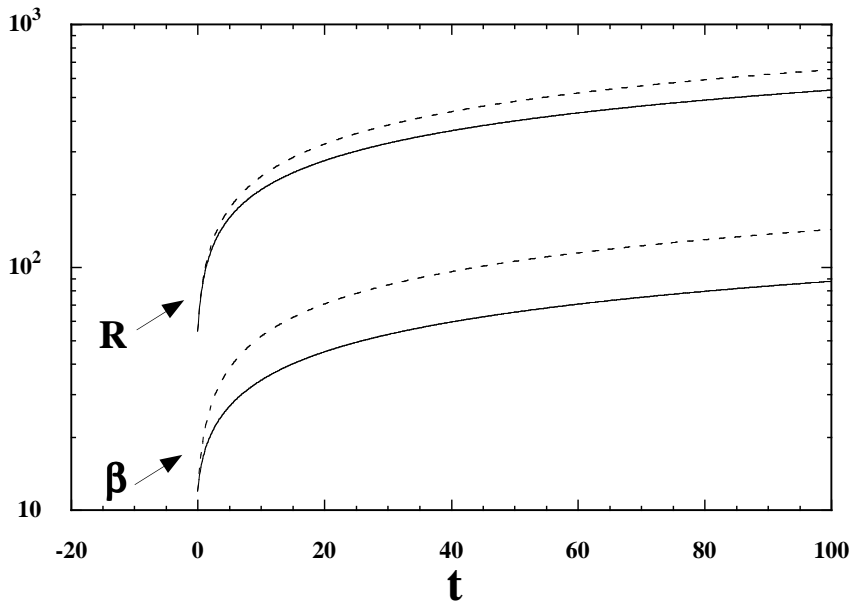}
\caption{The evolution of the large dimensions and $\beta$
that corresponds to the case (c) in fig.~\ref{small}.  The dotted curves
correspond to the case where the massive states are neglected (massless
states only).  }
\label{large}
\end{figure}
%%%

We have also made numerical simulations
for $r_0 > 1$.
When $\dot{\mu}_0>0$ and $\dot{\nu}_0>0$, both large and small dimensions
expand in the presence of positive pressures.  If $\dot{\nu}_0$ is largely
negative, the small dimensions contract by passing through $r=1$.
Meanwhile, if $|\dot{\nu}_0|\ll 1$, the small dimensions can exhibit bouncing
with $r>1$, instead of crossing $r=1$ [see the case (a) in
fig.~\ref{small}).  This means that the radius $r$ can grow in the presence of
the massive states.  Since the small dimensions continue to expand after
the bounce, this is not an ideal case where the small dimensions stay
small.  Nevertheless the small dimensions can be made small compared
to the large dimensions as long as $\dot{\mu}_0~\gsim~\dot{\nu}_0$.

When the massive states are taken into account,  this gives rise to
an extra source term for the energy $E$ in eq.~(\ref{b1}).  Then
$|\dot{\psi}_0|$ gets larger compared to the pure massless case.  Typically
this leads to the suppression of the r.h.s. of eq.~(\ref{b2}) via the
$e^{\psi}$ term, thereby yielding the smaller expansion rate $\dot{\mu}$
for the large dimensions.  The large
dimensions in the massive case grow slower relative to the massless case,
as seen in fig.~\ref{large}.  In addition, the massive effect suppresses
the growth of $\beta$, {\it i.e.} the temperature decreases faster in the
massless case.

As expected, the massive terms get smaller as the initial
value of $\beta$ is increased.  For example, in the cases shown in
fig.~\ref{small} and \ref{large}, the system is effectively described
by the massless states for $\beta_0>15$.
We also analyzed the behaviour of the system by varying the value $d$
with $1 \le d \le 8$.  We found that the numerical results are quite similar
to the case explained above ($d=4$).
As long as the conditions, $\dot{\mu}_0~\gsim~|\dot{\nu}_0|$ and 
$R_0~\gsim~r_0 \sim 1$,  are satisfied, the small dimensions are kept small, 
while the large ones expand as required in the presence of radiation and 
string matter.

We have also considered the case with the conservation of KK and/or winding
modes.  Under this circumstance, the partition function must be taken
summing only over the configurations that respect the conservation laws.
As before, this is done by introducing a chemical
potential corresponding to each conserved quantity in the partition function.
It turns out that no significant changes are found compared to the case
where no conservation laws are imposed.

%%%%%%%%%%%%%%%%%
\section{Conclusions}
%%%%%%%%%%%%%%%%%

We have studied string-gas cosmology at finite temperature in a
toroidal universe.  We make use of the dilation-gravity equations of
motion, satisfying the $R\rightarrow 1/R$ duality symmetry, to study the
evolution of the system, which is assumed adiabatic.  Our set up is as 
follows: the universe,
initially homogeneous, isotropic and in thermal equilibrium, is filled 
with an ideal gas of closed strings.  The 9 spatial dimensions,
compactified on a $9$-torus, evolve adiabatically starting from a Hagedorn
regime.

In order to find an acceptable equation of state that describes such a system 
in string theory at finite temperature, we evaluate the energy and
pressure in a microcanonical approach.  Close to the Hagedorn regime, the
scale factor $R$ exhibits a slow time evolution around $R\sim 1$, as shown
in fig.~\ref{hagedorn}.  In this case the dynamics of the system is
effectively described by a nearly constant energy and negligible pressure.
We found that the analytic solution in the Hagedorn regime shows very good
agreement with the full numerical result.  No substantial changes are
observed even when the conservation of Kaluza-Klein (KK) and winding modes
is imposed, as long as the conserved charges are of order one.

We have also investigated a ``low'' temperature regime, in which the equation of
state is derived in a canonical context.  We first considered the dynamics
of 3 ``large'' and 6 ``small'' compact dimensions in the presence of a pure
gas of radiation (given by the massless states).  It turns out, as
expected, that there exist interesting cosmological solutions where the
large dimensions continue to expand while the small dimensions
remain nearly constant and small relative to the large ones (see 
fig.~\ref{mlessevo}).  The attractor solutions for the large dimensions can 
be described by the evolution of the standard radiation dominant phase in 
FRW cosmology whereas the small dimensions always asymptotically approach 
to a constant value.  We also found bouncing solutions for the large 
dimensions if their Hubble rates are negative initially (see 
fig.~\ref{bounce}).

We then analysed the case where the massive string states are 
taken into account in
addition to the pure radiation.  The presence of the massive states typically  
leads to a slower expansion of the ``large'' dimensions relative to the 
massless case (see fig.~\ref{large}).  Meanwhile the behaviour of the small 
dimensions strongly depends on the initial conditions for $r$ and 
$\dot{r}$, resulting in either expansion or contraction of the small 
dimensions (see fig.~\ref{small}). 
The radius $r$ can be kept small as long as $r$ is initially close to unity,
since the pressure vanishes at the duality symmetric radius ($r=1$). 
The vanishing of the pressure at $r=1$ is a purely stringy effect, since it is due
to winding modes, whose negative contribution compensates that of KK states.
The important point is that, even in the presence of the massive state, there 
exist a wide range of the initial condition space for which the small 
dimensions are stabilized around the self-dual radius and are kept small 
relative to the large ones.  These behaviours are found to be insensitive 
to the number of large dimensions, $d$.  We also considered the case for 
the conservation of KK and winding modes and found no substantial change 
compared to the case without imposing the conservation laws.

In this work we did not investigate the later stage of the cosmological
evolution.  From the view point of cosmology, it is important to have an
inflationary stage in order to solve the major cosmological problems
(such as the flatness problem) as well as to generate the seeds for
large-scale structure.  One way to address this problem is to assume that
the dilaton (or moduli) acquires mass,
which may lead to inflation at later stages.
Recently, for example, an interesting proposal was made by Parry and
Steer \cite{Parry:2001zg}, who showed that inflation can occur
on a moving 3-brane due to the nonminimal coupling  of the dilaton
to the brane matter.  It was also shown in ref.~\cite{AFK} that brane 
inflation could result from the negative pressure of winding modes.  
Although it is not clear at present whether these proposals are directly 
applicable to our string-gas work, it is certainly of interest to extend 
the present analysis in that direction.

%%%%%%%%%%%%%%%%%%%%%%%%%%%%%%%%%%%%%%%%%%%%%%%
\vskip 25pt
\noindent
{\Large \bf Acknowledgements}
\vskip 10pt

We thank Robert Brandenberger, Daisuke Ida, Ian Kogan, Yann Mambrini, and particularly
Miguel Vazquez-Mozo, for very useful discussions and comments.  M.B.  
thanks CONACyT (Mexico).  The research of S.T. is financially supported 
from JSPS (No.  04942).  B.B and S.T. are grateful to SISSA for 
their kind hospitality and support while much of the work of this project
was completed.\\
This work was partially supported by the EC through the RTN network
``The quantum structure of space-time and the
geometric nature of fundamental interactions'', contract HPRN-CT-2000-00131.

%%%%%%%%%%%%%%%%%%%%%%%%%%%%%%%%%%%%%%%

%%%%%%%%%%%%%%%%%%%%%%%%%%%%%%%%%%%%%
\section*{APPENDIX : The one-loop string partition function}

A thermal canonical partition function at temperature $T$ can be
computed in the Euclidean approach by compactifying the time direction
on a circle of length $\beta=1/T$. We are then led to consider Type IIA/B string theory
with all ten space-time directions compactified.
For simplicity, we take the tori to be all rectangular ({\it i.e.}
simple products of circles) and compute the free energy for
the case in which $(9-d)$ spatial dimensions are compactified on small radii all equal
to a common value denoted $r$, whereas the remaining $d$ directions are taken very large,
and all equal to $R$. The free energy, after having unfolded the fundamental
region of the world-sheet torus, takes the following form (see {\it e.g.} \cite{AO_Tan}):
\bea F^{(d)}(\beta)=
-\frac{V_d}{2\pi\sqrt{\al}}\int_{-{1\over 2}}^{1\over
2}d\tau_1\int_0^{\infty}{d\tau_2\over\tau_2^{(3+d)/2}}\,
[\Lambda(r;\tau )]^{9-d}  \sum_{p=1}^\infty
e^{-{\beta^2p^2\over 4\pi\tau_2}}\, |M_2|^2(\tau )\,,
\label{free}
\eea
where
\be
\Lambda(r;\tau)= \sum_{m,n=-\infty}^{+\infty} q^{\frac{1}{4}( {m\over
r} + nr)^2}{\bar q}^{\frac{\al}{4}( {m\over r} - nr )^2}\,,
\label{Rlattice}
\ee
represents the contributions of the whole KK and winding
modes along the small dimensions, $\tau$ is the so-called modular 
parameter of the
world-sheet torus and $q=\exp(2i\pi \tau)$. In deriving eq.(\ref{free}) 
the winding
modes along the large dimensions have been completely neglected and the 
sum over the KK
modes has been approximated by an integral over continuous momenta.
Note that $p$ in eq.(\ref{free}) runs only over positive {\it odd} numbers
and correspond to taking the correct quantum statistic for bosons and fermions.
Taking only the term $p=1$ in the above sum corresponds to replacing
the quantum bosonic/fermionic distribution with 
the classical Maxwell-Boltzmann distribution.
$V_d$ is the volume of the large dimensions in $(4\pi^2)$ units\footnote{Recall that
we are setting $\al=1$.}:
\be
V_d \equiv  \frac{1}{(4\pi^2)^{d/2}} (2\pi R)^d=R^d\,,
\ee
The $M_2$ factor in eq.~(\ref{free}) encodes the contribution to 
the free energy of the whole
tower of massive strings, and can be expanded in powers of $q$:
\beq
M_2(\tau) = \frac{\theta_2(\tau)^4}{\eta(\tau)^{12}} =
\sum_{N=0}^{\infty} D(N)\,q^N\,.
\label{M2t}
\eeq
Here $\theta_2$ and $\eta$ are modular functions on the torus
(see {\it e.g.} \cite{PolBook} for an
explicit expression) and $D(N)$ is the degeneracy factor at level
$N$ ($D(0) = 16$, for example).
The value of $N$ corresponds to each string mass level.
The $\tau_1$ and $\tau_2$ integrals in eq.~(\ref{free}) can be easily
performed.  It is convenient to consider the term with $N=\bar N=m_i=n_i=0$
in eq.~(\ref{free}) separately from the remaining ones.  This is the
contribution of the purely massless states, which we will henceforth
denote as $F_{rad}$ (where ``rad'' stands for radiation).  We get
\be
F_{rad}^{(d)} = - \frac{R^d}{2\pi} D(0)^2 \Gamma
\left(\frac{d+1}{2}\right)(4\pi)^{(d+1)/2} \zeta(d+1) (1-2^{-(d+1)})
\beta^{-d-1}\,,
\label{freed0}
\ee
where $\xi(x)$ is the Riemann zeta-function. The remaining ``matter'' terms
give
\bea
F_{mat}^{(d)}= -\frac{V_d}{\pi}\sum_{m_i,\, n_i}\sum_{p=1}^\infty
\sum_{N,\, \bar N}
\bigg(\frac{2\pi M}{\beta p}\bigg)^{(d+1)/2} \delta_{m_in_i + N-\bar N,\,
0}\, D(N) D(\bar N) \, K_{(d+1)/2}(\beta p M)\, ,
\label{Fd}
\eea
where $K_{(d+1)/2}$ are modified Bessel functions, and
\beq
M= M(m_i,n_i,r_i,N,\bar N)
\equiv \sqrt{\sum_{i=1}^{9-d}\left ({m_i^2\over r^2}
+ n_i^2r^2 \right )+2(N+\bar N) }\,.
\label{M}
\eeq
The total free energy is given by
 \be
 F^{(d)} = F^{(d)}_{rad} \, + \, F^{(d)}_{mat}\,.
 \ee
The infinite sum over $N$ and $\bar N$ is not always convergent.
In fact, the degeneracy factors $D(N)$, for large values of $N$,
have a leading exponential behavior $D(N)\sim \exp(2\pi  \sqrt{2N})$.
On the other hand, for large values of its argument,
the modified Bessel function $K_n(z)$ admits an asymptotic expansion whose
leading term is $\sim \exp(-z)$. Hence, the sum over $N$ and $\bar N$
in eq.~(\ref{Fd}) converges only for $\beta> \beta_H=2\pi \sqrt{2}$.  The
temperature $T_H = 1/\beta_H$ is the Hagedorn temperature.

As long as we deal with a range of energies where eq.~(\ref{free})
converges and no large energy fluctuations are present,
we can work with the canonical ensemble. On the other hand,
for high energy densities a microcanonical description has to be used.
In this case the energy density of states is governed by the analytic structure
of the canonical partition function $Z=\exp(-\beta F)$, in the complex
$\beta$-plane. Taking into account the leading singularities of (\ref{free}), we can
parametrize the partition function as
\be
Z(\beta,R)  \simeq \frac{e^{\Lambda(\beta,R)}}
{\beta-\beta_H}\bigg(\frac{\eta_K}{\beta-\beta_K}
\bigg)^{18}\bigg(\frac{\eta_W}{\beta-\beta_W} \bigg)^{18}\,,
\label{ZHage}
\ee
where $\eta_K$ and $\eta_W$ are defined by eq.~(\ref{etas}) with
$\beta_K=\beta_H - \eta_K$ and $\beta_W=\beta_H - \eta_W$.
$\Lambda(\beta, R)$ is an entire function in $\beta$.
The microcanonical
energy distribution function $\Omega (E)$, is then given by
\be
\Omega (E)=\int_{\beta_H-i\infty}^{\beta_H+i\infty}\frac{d\beta}{2\pi i}\,Z(\beta,R)\,e^{\beta E}
\simeq \sum_{i=H,K,W}\oint_{C_i}\frac{d\beta}{2\pi i}\,Z(\beta,R)\,e^{\beta E},
\label{omega}
\ee
where $C_H$, $C_K$ and $C_W$ are the three contours encircling respectively the poles
in $\beta_H$, $\beta_K$ and $\beta_W$ in the complex $\beta$-plane.
The entropy and the rest of the thermodynamical quantities easily
follow from eq.~(\ref{omega}).

%%%%%%%%%%%%%%%%%


\begin{thebibliography}{99}
%%%%%%%%%%%%%%%%%

\bibitem{large}
I.~Antoniadis,
%``A Possible New Dimension At A Few Tev,''
Phys.\ Lett.\ B {\bf 246} (1990) 377; \\
I.~Antoniadis, C.~Munoz and M.~Quiros,
%``Dynamical supersymmetry breaking with a large internal dimension,''
Nucl.\ Phys.\ B {\bf 397} (1993) 515
[arXiv:hep-ph/9211309].

\bibitem{non_com}
R.~Brandenberger and P.~M.~Ho,
%``Noncommutative spacetime, stringy spacetime uncertainty
%  principle, and  density fluctuations,''
Phys.\ Rev.\ D {\bf 66} (2002) 023517
[AAPPS Bull.\  {\bf 12N1} (2002) 10]
[arXiv:hep-th/0203119]; \\
%%CITATION = HEP-TH 0203119;%%
S.~Alexander, R.~Brandenberger and J.~Magueijo,
%``Non-commutative inflation,''
arXiv:hep-th/0108190; \\
%%CITATION = HEP-TH 0108190;%%
S.~Alexander and J.~Magueijo,
%``Non-commutative geometry as a realization
%  of varying speed of light  cosmology,''
arXiv:hep-th/0104093.
%%CITATION = HEP-TH 0104093;%%

\bibitem{mald}
J.~Maldacena,
%``Non-Gaussian features of primordial fluctuations
%  in single field  inflationary models,''
arXiv:astro-ph/0210603.
%%CITATION = ASTRO-PH 0210603;%%

%\cite{Veneziano:1991ek}
\bibitem{Veneziano:1991ek}
G.~Veneziano,
%``Scale factor duality for classical and quantum strings,''
Phys.\ Lett.\ B {\bf 265}, 287 (1991); \\
%%CITATION = PHLTA,B265,287;%%
K.~A.~Meissner and G.~Veneziano,
%``Symmetries of cosmological superstring vacua,''
Phys.\ Lett.\ B {\bf 267}, 33 (1991).
%%CITATION = PHLTA,B267,33;%%

\bibitem{pbb}
M.~Gasperini and G.~Veneziano,
%``Pre - big bang in string cosmology,''
Astropart.\ Phys.\  {\bf 1} (1993) 317
[arXiv:hep-th/9211021].
%%CITATION = HEP-TH 9211021;%%

%\cite{Lidsey:1999mc}
\bibitem{Lidsey:1999mc}
J.~E.~Lidsey, D.~Wands and E.~J.~Copeland,
%``Superstring cosmology,''
Phys.\ Rept.\  {\bf 337}, 343 (2000)
[arXiv:hep-th/9909061].
%%CITATION = HEP-TH 9909061;%%

%\cite{Gasperini:2002}
\bibitem{Gasperini:2002}
M.~Gasperini and G.~Veneziano,
% ``The Pre-Big Bang Scenario in String Cosmology,''
arXiv:hep-th/0207130.
%%CITATION = HEP-TH 0207130%%

\bibitem{ekpyr}
J.~Khoury, B.~A.~Ovrut, P.~J.~Steinhardt and N.~Turok,
%``The ekpyrotic Universe: Colliding branes and
% the origin of the hot big  bang,''
Phys.\ Rev.\ D {\bf 64}, 123522 (2001)
[arXiv:hep-th/0103239];
%\cite{Khoury:2001zk}
%J.~Khoury, B.~A.~Ovrut, P.~J.~Steinhardt and N.~Turok,
%%``Density perturbations in the ekpyrotic scenario,''
Phys.\ Rev.\ D {\bf 66}, 046005 (2002)
[arXiv:hep-th/0109050].
%%CITATION = HEP-TH 0109050;%%
%N. D. Antunes, E. J. Copeland, M. Hindmarsh,
%A. Lukas, hep-th/0208219

%\cite{Steinhardt:2001vw}
\bibitem{cyclic}
P.~J.~Steinhardt and N.~Turok,
%``A cyclic model of the Universe,''
arXiv:hep-th/0111030;
%P.~J.~Steinhardt and N.~Turok,
%%``Cosmic evolution in a cyclic universe,''
Phys.\ Rev.\ D {\bf 65}, 126003 (2002)
[arXiv:hep-th/0111098].
%%CITATION = HEP-TH 0111098;%%

\bibitem{orbsing}
H.~Liu, G.~Moore and N.~Seiberg,
%``Strings in time-dependent orbifolds,''
JHEP {\bf 0210}, 031 (2002)
[arXiv:hep-th/0206182];
%%CITATION = HEP-TH 0206182;%%
%\cite{Elitzur:2002rt}
S.~Elitzur, A.~Giveon, D.~Kutasov and E.~Rabinovici,
%``From big bang to big crunch and beyond,''
JHEP {\bf 0206}, 017 (2002)
[arXiv:hep-th/0204189].
%%CITATION = HEP-TH 0204189;%%

\bibitem{BV}
R.~H.~Brandenberger and C.~Vafa,
%``Superstrings In The Early Universe,''
Nucl.\ Phys.\ B {\bf 316} (1989) 391.
%%CITATION = NUPHA,B316,391;%%

\bibitem{TV}
A.~A.~Tseytlin and C.~Vafa,
%``Elements of string cosmology,''
Nucl.\ Phys.\ B {\bf 372} (1992) 443
[arXiv:hep-th/9109048].
%%CITATION = HEP-TH 9109048;%%

\bibitem{Tse}
A.~A.~Tseytlin,
%``Dilaton, winding modes and cosmological solutions,''
Class.\ Quant.\ Grav.\  {\bf 9} (1992) 979
[arXiv:hep-th/9112004].
%%CITATION = HEP-TH 9112004;%%

\bibitem{Miguel}
M.~A.~Osorio and M.~A.~Vazquez-Mozo,
%``Variations on Kaluza-Klein cosmology,''
Mod.\ Phys.\ Lett.\ A {\bf 8} (1993) 3111
[arXiv:hep-th/9305137];
%%CITATION = HEP-TH 9305137;%%
%``String variations on Kaluza-Klein cosmology,''
Mod.\ Phys.\ Lett.\ A {\bf 8} (1993) 3215
[arXiv:hep-th/9305138].
%%CITATION = HEP-TH 9305138;%%

\bibitem{CR}
G.~B.~Cleaver and P.~J.~Rosenthal,
%``String cosmology and the dimension of space-time,''
Nucl.\ Phys.\ B {\bf 457} (1995) 621
[arXiv:hep-th/9402088].
%%CITATION = HEP-TH 9402088;%%

\bibitem{Sakella}
M.~Sakellariadou,
%``Numerical Experiments in String Cosmology,''
Nucl.\ Phys.\ B {\bf 468} (1996) 319
[arXiv:hep-th/9511075].
%%CITATION = HEP-TH 9511075;%%

\bibitem{Alex}
S.~Alexander, R.~H.~Brandenberger and D.~Easson,
%``Brane gases in the early universe,''
Phys.\ Rev.\ D {\bf 62} (2000) 103509
[arXiv:hep-th/0005212].
%%CITATION = HEP-TH 0005212;%%

\bibitem{Brand}
R.~Brandenberger, D.~A.~Easson and D.~Kimberly,
%``Loitering phase in brane gas cosmology,''
Nucl.\ Phys.\ B {\bf 623} (2002) 421
[arXiv:hep-th/0109165].
%%CITATION = HEP-TH 0109165;%%

\bibitem{Easson}
D.~A.~Easson,
%``Brane gases on K3 and Calabi-Yau manifolds,''
arXiv:hep-th/0110225.
%%CITATION = HEP-TH 0110225;%%

\bibitem{EGJ}
R.~Easther, B.~R.~Greene and M.~G.~Jackson,
%``Cosmological string gas on orbifolds,''
Phys.\ Rev.\ D {\bf 66} (2002) 023502
[arXiv:hep-th/0204099].
%%CITATION = HEP-TH 0204099;%%

\bibitem{Watson}
S.~Watson and R.~H.~Brandenberger,
%``Isotropization in brane gas cosmology,''
arXiv:hep-th/0207168.
%%CITATION = HEP-TH 0207168;%%

\bibitem{Easther}
R.~Easther, B.~R.~Greene, M.~G.~Jackson and D.~Kabat,
%``Brane gas cosmology in M-theory: Late time behavior,''
arXiv:hep-th/0211124.
%%CITATION = HEP-TH 0211124;%%

\bibitem{Leblanc}
Y.~Leblanc,
%``Cosmological Aspects Of The Heterotic String Above The Hagedorn
% Temperature,''
Phys.\ Rev.\ D {\bf 38} (1988) 3087.
%%CITATION = PHRVA,D38,3087;%%

\bibitem{DP1}
T.~Damour and A.~M.~Polyakov,
%``String theory and gravity,''
Gen.\ Rel.\ Grav.\  {\bf 26}, 1171 (1994)
[arXiv:gr-qc/9411069].
%%CITATION = GR-QC 9411069;%%

\bibitem{run}
T.~Damour, F.~Piazza and G.~Veneziano,
%``Violations of the equivalence principle in a
% dilaton-runaway scenario,''
Phys.\ Rev.\ D {\bf 66}, 046007 (2002)
[arXiv:hep-th/0205111];
%%CITATION = HEP-TH 0205111;%%
%T.~Damour, F.~Piazza and G.~Veneziano,
%%``Runaway dilaton and equivalence principle violations,''
Phys.\ Rev.\ Lett.\  {\bf 89}, 081601 (2002)
[arXiv:gr-qc/0204094].
%%CITATION = GR-QC 0204094;%%

\bibitem{RDSS}
S.~Randjbar-Daemi, A.~Salam and J.~Strathdee,
%``On Kaluza-Klein Cosmology,''
Phys.\ Lett.\ B {\bf 135} (1984) 388.
%%CITATION = PHLTA,B135,388;%%

%\cite{Alvarez-Gaume:1992re}
\bibitem{Alvarez-Gaume:1992re}
L.~Alvarez-Gaume and M.~A.~Vazquez-Mozo,
%``Topics in string theory and quantum gravity,''
arXiv:hep-th/9212006,
%%CITATION = HEP-TH 9212006;%%
section 4.3 and references therein.

\bibitem{Hage}
R. Hagedorn, Suppl. Nuovo Cimento {\bf 3} (1965) 147.

\bibitem{DJT}
N.~Deo, S.~Jain and C.~I.~Tan,
%``Strings At High-Energy Densities And Complex Temperature,''
Phys.\ Lett.\ B {\bf 220} (1989) 125.

\bibitem{DJT2}
N.~Deo, S.~Jain and C.~I.~Tan,
%\alpha``String Statistical Mechanics Above
% Hagedorn Energy Density,''
Phys.\ Rev.\ D {\bf 40} (1989) 2626.
%%CITATION = PHLTA,B220,125;%%

\bibitem{hawking}
S.~W.~Hawking,
%``Black Holes And Thermodynamics,''
Phys.\ Rev.\ D {\bf 13} (1976) 191.
%%CITATION = PHRVA,D13,191;%%

\bibitem{AEK}
M.~Axenides, S.~D.~Ellis and C.~Kounnas,
%``Universal Behavior Of D-Dimensional Superstring Models,''
Phys.\ Rev.\ D {\bf 37} (1988) 2964; \\
M.~J.~Bowick and L.~C.~Wijewardhana,
%``Superstrings At High Temperature,''
Phys.\ Rev.\ Lett.\  {\bf 54} (1985) 2485.
%%CITATION = PRLTA,54,2485;%%

\bibitem{AW}
J.~J.~Atick and E.~Witten,
%''The Hagedorn Transition and the Number of Degrees of
%Freedom of String Theory,''
Nucl.\ Phys. \ B {\bf 310} (1988) 291.

%\cite{Gasperini:1996fu}
\bibitem{Gasperini:1996fu}
M.~Gasperini, M.~Maggiore and G.~Veneziano,
%``Towards a non-singular pre-big bang cosmology,''
Nucl.\ Phys.\ B {\bf 494}, 315 (1997)
[arXiv:hep-th/9611039]; \\
R.~Brustein and R.~Madden,
Phys.\ Rev.\ D {\bf 57}, 712 (1998)
[arXiv:hep-th/9708046].
%%CITATION = HEP-TH 9708046;%%

%\cite{Tsujikawa:2002qc}
\bibitem{Tsujikawa:2002qc}
S.~Tsujikawa, R.~Brandenberger and F.~Finelli,
%``On the construction of nonsingular pre-big-bang
% and ekpyrotic  cosmologies and the resulting
% density perturbations,''
Phys.\ Rev.\ D {\bf 66}, 083513 (2002)
[arXiv:hep-th/0207228].
%%CITATION = HEP-TH 0207228;%%

%\cite{Parry:2001zg}
\bibitem{Parry:2001zg}
M.~F.~Parry and D.~A.~Steer,
%``Brane gas inflation,''
JHEP {\bf 0202}, 032 (2002)
[arXiv:hep-ph/0109207].
%%CITATION = HEP-PH 0109207;%%

\bibitem{AFK}
S.~Abel, K.~Freese and I.~I.~Kogan,
JHEP {\bf 0101}, 039 (2001)
% Hagedorn inflation of D-branes.
[archiv:hep-th/0005028];
%``Hagedorn inflation: Open strings on branes 
% can drive inflation,''
arXiv:hep-th/0205317.
%%CITATION = HEP-TH 0205317;%%

\bibitem{AO_Tan}
E.~Alvarez and M.~A.~Osorio,
%``Superstrings At Finite Temperature,''
Phys.\ Rev.\ D {\bf 36}, 1175 (1987); \\
%%CITATION = PHRVA,D36,1175;%%
K.~H.~O'Brien and C.~I.~Tan,
%``Modular Invariance Of Thermopartition Function
% And Global Phase Structure Of Heterotic String,''
Phys.\ Rev.\ D {\bf 36} (1987) 1184.
%%CITATION = PHRVA,D36,1184;%%

\bibitem{PolBook}
J.  Polchinski, {\it String Theory}, Volume 1, Cambridge, page. 214-216.

%\cite{Deo:1991af}
%\bibitem{Deo:1991af}
%N.~Deo, S.~Jain and C.~I.~Tan,
%%``The ideal gas of strings,''
%HUTP-90-A079
%%\href{http://www.slac.stanford.edu/spires/find/hep/www?r=hutp-90-a079}
%%{SPIRES entry}
%{\it Based on talk given at Int. Colloq. on Modern Quantum Field Theory,
%Bombay, India, Jan 8-14, 1990}

%%\cite{Leblanc:1988eq}
%\bibitem{Leblanc:1988eq}
%Y.~Leblanc,
%%``Cosmological Aspects Of The Heterotic String
%% Above The Hagedorn Temperature,''
%Phys.\ Rev.\ D {\bf 38} (1988) 3087.
%%%CITATION = PHRVA,D38,3087;%%


\end{thebibliography}
\end{document}